\documentclass[twocolumn,prb,showpacs,preprintnumbers,amsmath,amssymb]{revtex4}
\usepackage{graphicx}
\usepackage{dcolumn}
\usepackage{bm}
\usepackage{epsfig}

\begin{document}

\title{Neutron scattering study of a quasi-2D spin-1/2 dimer system Piperazinium Hexachlorodicuprate under hydrostatic pressure}

\author{Tao Hong}
\affiliation{Neutron Scattering Sciences Division, Oak Ridge National Laboratory, Oak Ridge, Tennessee 37831-6393, USA}
\author{C. Stock}
\author{I. Cabrera}
\author{C. Broholm}
\affiliation{Department of Physics and Astronomy, The Johns Hopkins University, Baltimore,
Maryland 21218, USA}
\affiliation{ NIST Center for Neutron Research, National Institute
of Standards and Technology, Gaithersburg, MD 20899, USA.}
\author{Y. Qiu}
\affiliation{ NIST Center for Neutron Research, National Institute
of Standards and Technology, Gaithersburg, MD 20899, USA.}
\affiliation{ Department of Materials Science and Engineering,
University of Maryland, College Park, MD 20742, USA.}
\author{J. B. Leao}
\author{S. J. Poulton}
\author{J. R. D. Copley}
\affiliation{ NIST Center for Neutron Research, National Institute
of Standards and Technology, Gaithersburg, MD 20899, USA.}

\date{\today}

\begin{abstract}
We report inelastic neutron scattering study of a quasi-two-dimensional \emph{S}=1/2 dimer system Piperazinium Hexachlorodicuprate under hydrostatic pressure. The spin gap $\Delta$ becomes softened with the increase of the hydrostatic pressure up to $P=9.0$ kbar. The observed threefold degenerate triplet excitation at $P=6.0$ kbar is consistent with the theoretical prediction and the bandwidth of the dispersion relation is unaffected within the experimental uncertainty. At $P=9.0$ kbar the spin gap is reduced to $\Delta$=0.55~meV from $\Delta$=1.0~meV at ambient pressure.
\end{abstract}

\pacs{75.10.Jm, 75.50.Ee}

\maketitle

\section{Introduction}
Quantum phase transition has been a long studied fundamental issue to understand the universality of quantum critical behavior in many-body systems.\cite{Sach99} Gapped Heisenberg spin-1/2 dimer systems have the potential to exhibit quantum critical phenomena in their excitation spectra as a function of applied magnetic field or hydrostatic pressure. In the past decades, much attention has been focused on the quantum phase transition at which the spin gap is closed by an applied magnetic field. In the vicinity of this transition, the $S_z=1$ excitations above the spin gap behave like canonical bosons, and the transition maps simply to the Bose-Einstein condensation (BEC) of a dilute Bose gas. Such BEC has been extensively studied both theoretically\cite{Affl91:43,Giama99:59,Thier08:4} and experimentally.\cite{Ruegg03:423,Zapf06:96,Stone06:96,Garlea07:98}

Whereas under a hydrostatic pressure, it has been observed that the spin gap $\Delta$ is reduced in a one-dimensional \emph{S}=1/2 quantum spin ladder material IPA-CuCl$_3$.\cite{Hong08:78} In the three-dimensional \emph{S}=1/2 dimer systems $\rm TlCuCl_3$\cite{Oosa03:72,Oosa04:73,Goto04:73,Rueg04:93,Rueg08:100} and $\rm KCuCl_3$\cite{Goto06:75}, the spin gap collapses above a certain critical pressure $P_c$. Hence these systems transition from a gapped singlet state to an ordered antiferromagnetic state under the effect of hydrostatic pressure. The difference between the field and pressure-induced quantum phase transitions is that the former arises from softening of one of the three members of a triplet, while the latter transition arises from softening of all three modes, which are degenerate below $P_c$.\cite{Mats04:69}

Thus far, no experimental work on such a two-dimensional (2D) spin-1/2 dimer system under a hydrostatic pressure has been done. Recently, Stone \emph{et al.} reported the bulk and inelastic neutron scattering (INS) measurements study of a frustrated quasi 2D spin-1/2 dimer system---Piperazinium Hexachlorodicuprate (PHCC).\cite{stone01:64} The crystal structure of PHCC is composed of Cu-Cl sheets that span the {\bf \emph{a}}-{\bf \emph{c}} plane and are separated by layers of piperazinium molecules. The in plane magnetic interactions are much stronger than the interplane interactions. This makes PHCC an excellent physical realization of a 2D quantum antiferromagnet. The magnetic excitations at zero field are dominated by a dispersive triplon with a bandwidth of 1.7 meV and a spin gap $\Delta\simeq$1.0 meV in the $(h,0,\ell)$ plane.\cite{stone01:64} This makes PHCC a good candidate to study the quantum critical phenomena of a 2D quantum spin-1/2 dimer system under hydrostatic pressure. In this paper, we explore the hydrostatic pressure effect in PHCC. Either the spin gap or the magnetic excitation spectrum were measured in PHCC under hydrostatic pressure up to 9.0 kbar. We observed the degenerate triplet spectrum and a softening of the spin gap $\Delta$ with increasing pressure.

\section{Sample and neutron instrumentation}
Usually, for a high pressure experiment, sample space is limited and neutron beam is attenuated due to the thick wall of the pressure cell, which makes INS measurements hard to carry out.

The single crystalline samples of PHCC were prepared using the same method as described in Ref.~\onlinecite{Batt88:2}. INS measurements were performed using the cold neutron triple axis spectrometer SPINS and the disk chopper time-of-flight spectrometer DCS\cite{Copley03} at the NIST Center for Neutron Research. The sample used for SPINS consisted of two single crystals with a total mass of 0.3 g and a 1.0$^\circ$ mosaic spread. The sample used for DCS consisted of four single crystals with a total mass of 1.0 g coaligned within 3.0$^\circ$. The SPINS measurements were made with a fixed final energy $E_f$=3.7 meV and horizontal beam divergences given by $^{58}$Ni guide-open-80'-open collimations. PG and BeO filters were placed before and after the sample, respectively, to remove higher order beam contamination. DCS measurements were made with the incident neutron wavelength fixed at $\lambda=4.8$ $\AA$, covering the sample rotational angles with a range of 48 degrees and probing transferred energies up to 2.3 meV. The sample with holder was mounted in an aluminum (at SPINS) or a stainless steel (at DCS) pressure cell and inserted in a standard 4He cryostat. The maximum pressure for the aluminum cell is 6 kbar while the steel cell is capable of 10 kbar. The pressure cell also contained PG (002) platelet for the pressure calibration.\cite{Hanf89:39} The pressure transducing medium was helium gas. In all measurements, sample was oriented with its reciprocal $(h,0,\ell)$ plane in the horizontal plane. Wave-vector transfer is indexed as $\textbf{\emph{Q}}=h{\bf \emph{a}}^\ast+\ell{\bf \emph{c}}^\ast$.

\section{Data analysis and discussion}
\begin{figure}
\includegraphics[width=7.5cm,bbllx=70,bblly=20,bburx=545,
  bbury=750,angle=0,clip=]{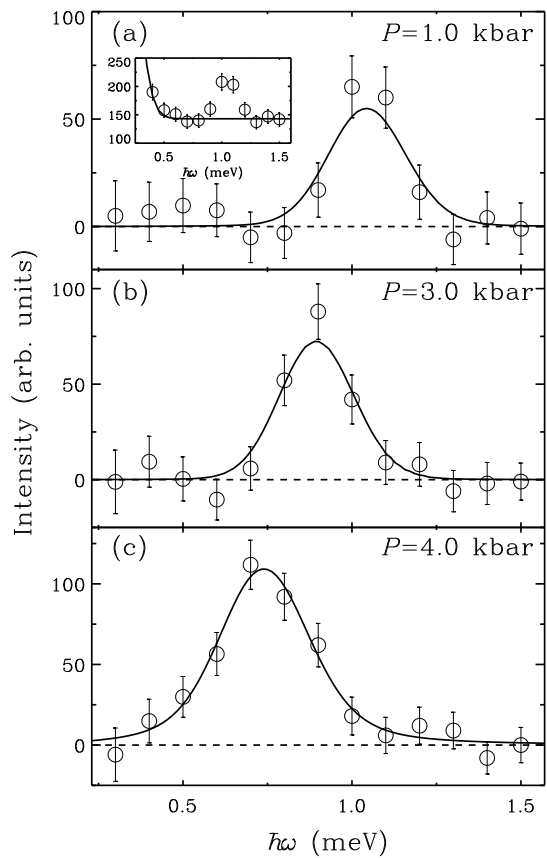}
\caption{Background-subtracted constant-\textbf{Q}=(0.5,0,-1.5) scans measured at SPINS in PHCC
at \emph{T}=2.3 K (a) \emph{P}=1 kbar, (b) \emph{P}=3 kbar, and (c) \emph{P}=4 kbar. Solid lines are fits to
the model as described in the text after convolution with the instrumental resolution function. Dashed lines indicate the level of zero. Inset: Raw data of constant-\textbf{Q}=(0.5,0,-1.5) scan measured at \emph{P}=1 kbar. The
solid line is a Gaussian and a linear term fit to account for the background contribution. Throughout error bars indicate plus minus the standard deviation, $\sigma$.}
\label{fig1}
\end{figure}

Figure~\ref{fig1} shows the background-subtracted data collected from SPINS at the antiferromagnetic (AFM) zone center \textbf{\emph{Q}}=(0.5,0,-1.5) and \emph{T}$\simeq$2.3 K with the hydrostatic pressure up to \emph{P}=4.0 kbar. The background (shown as a solid line in inset of Fig.~\ref{fig1}(a)) was determined at the same hydrostatic pressure by making an energy scan at \textbf{\emph{Q}}\,=\,(0.4,0,-1.5), away from the magnetic zone center, with the same instrument configuration and by fitting the results to a Gaussian profile, plus a term linear in energy, over the range where no magnetic excitation is expected. The location of singlet-triplet spin gap was determined by least-squares fitting to the following scattering function satisfying a detailed balance condition and numerically convolved with the calculated instrumental resolution function. We used the same two-Lorentzian Damped Harmonic Oscillator response function for PHCC, as previously applied to the study of finite-temperature dependent energy spectra.\cite{Stone06:440} At each pressure, the fitting parameters of this model include the spin gap $\Delta$, the intrinsic excitation width $\Gamma$, and an overall intensity prefactor A:
 \begin{eqnarray}
 S(\textbf{Q},\omega)& = &
 \frac{A}{1-exp(-\beta\hbar\omega)}\left[  \frac{\Gamma}{(\hbar\omega-\epsilon_{\bf Q})^2+\Gamma^2} \right. \nonumber\\
  & + & \left. \frac{\Gamma}{(\hbar\omega+\epsilon_{\bf Q})^2+\Gamma^2}
 \right].
 \label{sqw}
 \end{eqnarray}

The data agree very well with the model in the entire scan range. Excitation peaks at all pressures are resolution-limited. The results are plotted in solid lines in Figs.~\ref{fig1}(a), (b), and (c).

While a conventional triple-axis spectrometer is well suited to the study of spin gap excitation in PHCC, a time-of-flight instrument can be used to explore rather large regions in (\textbf{Q}, $\omega$) phase space because many detectors simultaneously collect neutrons over a wide range of scattered energies.

\begin{figure}
\includegraphics[width=7.5cm,bbllx=115,bblly=370,bburx=550,bbury=725,angle=0,clip=]{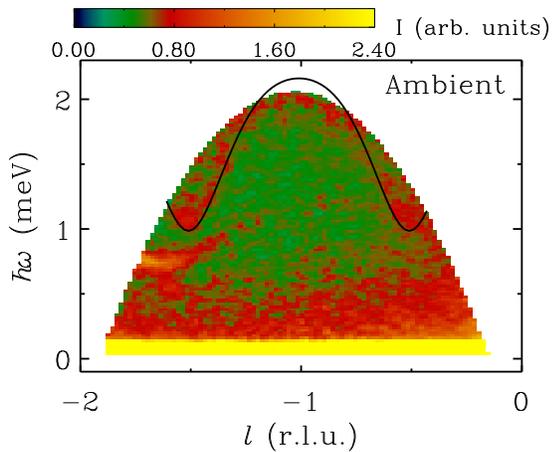}
\caption{(Color online) The single crystalline inelastic neutron scattering intensity along the reciprocal (0.5,0,$\ell$) direction measured at DCS for PHCC at \emph{T}=1.5 K and ambient pressure. Solid line is the one-triplon dispersion relation for PHCC at ambient pressure as determined from Ref. \onlinecite{Stone06:440}. The figure was obtained by averaging the data in bins of size d$\hbar\omega$=0.02 meV and d$\ell$=0.02 (r.l.u.).}
\label{fig2}
\end{figure}

Figure~\ref{fig2} shows the INS intensity in arbitrary units measured at DCS as a function of transferred energy $\hbar\omega$ and \textbf{Q}=(0.5,0,$\ell$) at \emph{T}=1.5 K and ambient pressure.\cite{Azuah09:114} The solid line indicates the magnetic one-triplon dispersion relation at ambient pressure for PHCC.\cite{stone01:64} Note that the dispersion relation is consistent with the observed intensity maxima, confirming that the experiment was able to probe magnetic scattering from the small PHCC sample. Because there was still considerable amount of Helium inside the pressure cell which condenses to solid at base temperature and high pressure, the observed scattering intensity also includes a significant contribution from excitation of roton in solid Helium near $\hbar\omega\simeq$0.75 meV.\cite{Fak98:112} Fortunately, the form of solid Helium is usually polycrystalline, therefore the excitation spectrum in solid Helium should be quite isotropic. Since the magnetic excitation in PHCC along the reciprocal (0,0,$\ell$) direction is almost dispersionless around 2.7 meV,\cite{stone01:64} which is beyond our experimentally accessible range, it allows us to treat the scattering intensity along the reciprocal (0,0,$\ell$) direction as background.

\begin{figure}[h]
\includegraphics[width=7.8cm,bbllx=115,bblly=120,bburx=530,bbury=695,angle=0,clip=]{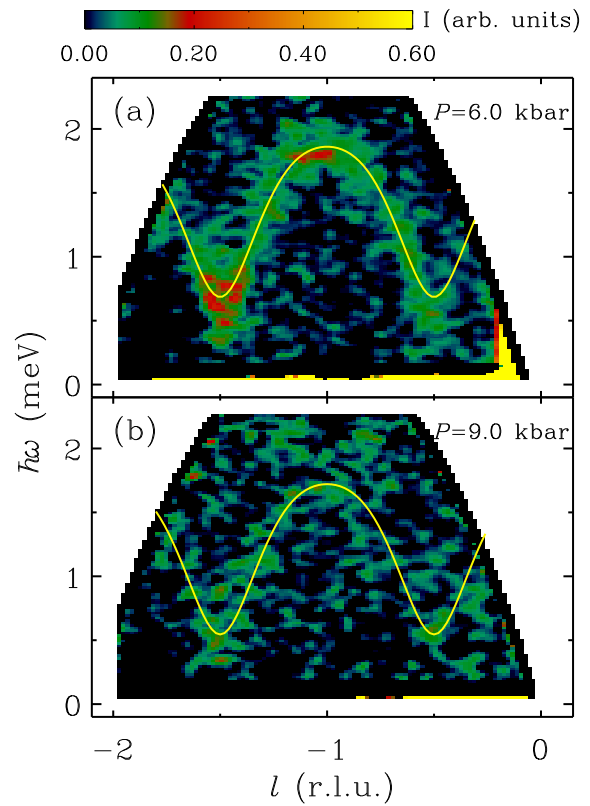}
\caption{(Color online) Background-subtracted inelastic neutron scattering intensity along the reciprocal (0.5,0,$\ell$) direction measured at DCS for PHCC at \emph{T}=1.5 K, and (a) \emph{P}=6.0 kbar and (b) \emph{P}=9.0 kbar. The background from solid helium excitation was subtracted as discussed in the text. Solid lines are the one-triplon dispersion relation for PHCC at ambient pressure lowered by (a) 0.30 meV and (b) 0.45 meV. The figure was obtained by averaging the data in bins of size d$\hbar\omega$=0.04 meV and d$\ell$=0.04 (r.l.u.). }
\label{fig3}
\end{figure}

\begin{figure}[t]
\includegraphics[width=7.5cm,bbllx=95,bblly=120,bburx=535,bbury=645,angle=0,clip=]{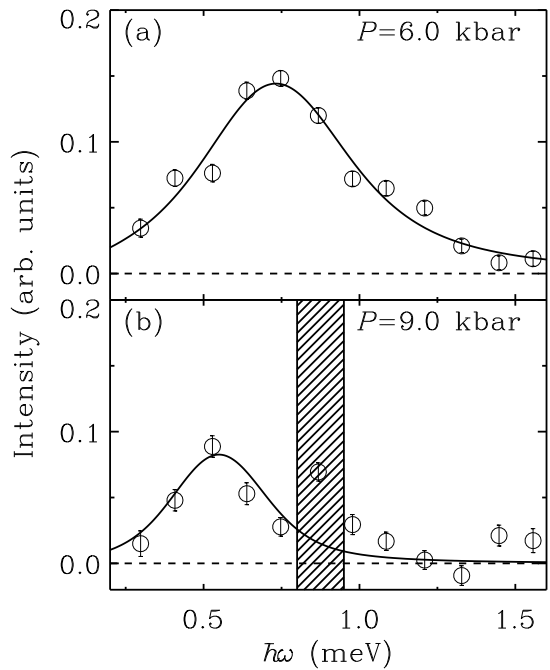}
\caption{Energy dependence of the magnetic scattering intensity for PHCC averaged over the range of -1.6$<\ell<$-1.4 from Fig.~\ref{fig3} at \emph{T}=1.5 K, and (a) \emph{P}=6.0 kbar and (b) \emph{P}=9.0 kbar. Solid lines are the fits to Eq.~1 convolved with the instrumental resolution function. Dashed lines indicate the level of zero. The shaded area is excluded due to a contamination by the solid helium excitation spectrum. Throughout error bars indicate plus minus the standard deviation, $\sigma$.}
\label{fig4}
\end{figure}

\begin{figure}[t]
\includegraphics[width=7.5cm,bbllx=95,bblly=370,bburx=535,bbury=690,angle=0,clip=]{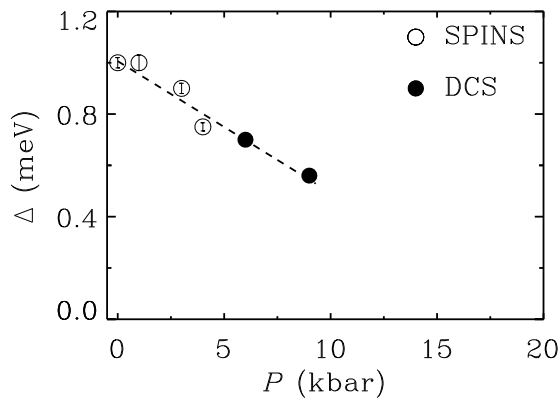}
\caption{The pressure-dependence of spin gap $\Delta$ in PHCC. The spin gap at ambient pressure was reproduced from Ref.~\onlinecite{stone01:64}. The dashed line is a guide to the eye.}
\label{fig5}
\end{figure}

Figure~\ref{fig3} shows the magnetic scattering intensity at $P$=6.0 and 9.0 kbar after subtracting such background contribution from the solid helium. Clearly, the whole triplet excitation spectrum  at $P$=6.0 kbar in PHCC is shifted together towards a lower energy and the dispersion bandwidth is estimated to be 1.10(15) meV,\cite{endnote} which is same as 1.18 meV at ambient pressure. To precisely determine $\Delta$ value, the magnetic scattering intensity from Fig.~\ref{fig3} was averaged over the range of -1.6$<\ell<$-1.4, plotted as a function of $\hbar\omega$ as shown in Fig.~\ref{fig4}, and then fitted to Eq.~\ref{sqw} after being convoluted with the instrumental resolution function. The measured spin gap $\Delta$ as a function of applied hydrostatic pressure is summarized in Fig.~\ref{fig5}.

However, the magnetic intensity at \emph{P}=9.0 kbar is much weaker and the excitation spectrum becomes blurred as shown in Fig.~\ref{fig3}(b). It could be caused by the less successful subtraction of the non-magnetic background or the variation in the exchange interactions. In the latter case, a complete understanding of the changes of the exchange interactions caused by an applied hydrostatic pressure would require a detailed structural investigation of the alterations to bond length and angles. For PHCC, the structure is complex due to geometrically frustrated interactions and at least 8 exchange interactions need to be considered.\cite{stone01:64} Such a study lies beyond the scope of the present analysis.

\section{CONCLUSION}
In summary, we performed inelastic neutron scattering experiments to measure the magnetic excitation spectrum of the quasi-2D spin-1/2 dimer system PHCC under hydrostatic pressure. Both SPINS and DCS experiments showed the softening of the energy gap with increasing hydrostatic pressure up to \emph{P}=9.0 kbar. The driving mechanism of this behavior is the variation in strength of exchange interactions as a function of hydrostatic pressure. If a pressure cell with higher limit is developed, future work will focus on the determination of possible magnetic order and excitations in the high pressure phase. It will also be important to determine whether the system remains quasi-two-dimensional above 9 kbar.

\begin{acknowledgments}
We thank R. Paul for help with neutron activation analysis. The DAVE program is supported by the NSF under Agreement No. DMR-0454672. The work at ORNL was partially funded by the Division of Scientific User Facilities, Office of BES, DOE. The work at JHU was supported by the NSF through Grants No. DMR-0306940 and No. DMR-0706553. The work at NIST utilized facilities supported in part by the NSF under Agreement No. DMR-0454672.
\end{acknowledgments}

\end{document}